\documentclass[preprint2]{aastex}

\shorttitle{Chemistry in Prestellar Cores}
\shortauthors{Aikawa et al.}

\received{2003 January 29}
\begin{document}

%% LaTeX will automatically break titles if they run longer than
%% one line. However, you may use \\ to force a line break if
%% you desire.

\title{Molecular Evolution in Collapsing Prestellar Cores II: The
Effect of Grain-surface Reactions}

%% Use \author, \affil, and the \and command to format
%% author and affiliation information.
%% Note that \email has replaced the old \authoremail command
%% from AASTeX v4.0. You can use \email to mark an email address
%% anywhere in the paper, not just in the front matter.
%% As in the title, you can use \\ to force line breaks.

\author{Yuri Aikawa}
\affil{Department of Earth and Planetary Sciences, Kobe University,
Kobe 657-8501, Japan}

\author{Nagayoshi Ohashi}
\affil{Academia Sinica Institute of Astronomy and Astrophysics,
P.O. Box 23-141, Taipei 106, Taiwan}

\and

\author{Eric Herbst}
\affil{Departments of Physics, Chemistry, and Astronomy, The Ohio 
State University,
Columbus, OH 43210, USA}

%% Mark off your abstract in the ``abstract'' environment. In the manuscript
%% style, abstract will output a Received/Accepted line after the
%% title and affiliation information. No date will appear since the author
%% does not have this information. The dates will be filled in by the
%% editorial office after submission.

\begin{abstract}
The molecular evolution that occurs in collapsing prestellar cores is
investigated.  To model the dynamics, we adopt the Larson-Penston
solution and analogues with slower rates of collapse.  For the
chemistry, we utilize the new standard model (NSM) with the addition
of deuterium fractionation and grain-surface reactions treated via the
modified rate approach.  The use of surface reactions distinguishes
the present work from our previous model.  We find that these
reactions efficiently produce H$_2$O, H$_2$CO, CH$_3$OH, N$_2$, and
NH$_3$ ices.  In addition, the surface chemistry influences the
gas-phase abundances in a variety of ways.  For example, formation of
molecular nitrogen on grain surfaces followed by desorption into the
gas enhances the abundance of this gas-phase species and its daughter
products N$_{2}$H$^{+}$ and NH$_{3}$.  The current reaction network
along with the Larson-Penston solution allows us to reproduce
satisfactorily most of the molecular column densities and their radial
distributions observed in L1544.  The agreement tends to worsen with
models that include strongly delayed collapse rates.  Inferred
radial distributions in terms of fractional abundances are somewhat
harder to reproduce. In addition to
our standard chemical model, we have also run a model with the UMIST
gas-phase chemical network.  The abundances of gas-phase
sulphur-bearing molecules such as CS and CCS are significantly
affected by uncertainties in the gas-phase chemical network.  In all
of our models, the column density of N$_2$H$^+$ monotonically
increases as the central density of the core increases during collapse
from $3\times 10^4$ cm$^{-3}$ to $3\times 10^7$ cm$^{-3}$.  Thus, the
abundance of this ion can be a probe of evolutionary stage.  Molecular
D/H ratios in assorted cores
%in the Taurus molecular cloud
are best
reproduced in the Larson-Penston picture with the conventional rate
coefficients for fractionation reactions.  If we adopt the newly
measured and calculated rate coefficients, the D/H ratios, especially
N$_2$D$^+$/N$_2$H$^+$, become significantly lower than the observed
values.
\end{abstract}

%% Keywords should appear after the \end{abstract} command. The uncommented
%% example has been keyed in ApJ style. See the instructions to authors
%% for the journal to which you are submitting your paper to determine
%% what keyword punctuation is appropriate.


%% From the front matter, we move on to the body of the paper.
%% In the first two sections, notice the use of the natbib \citep
%% and \citet commands to identify citations.  The citations are
%% tied to the reference list via symbolic KEYs. The KEY corresponds
%% to the KEY in the \bibitem in the reference list below. We have
%% chosen the first three characters of the first author's name plus
%% the last two numeral of the year of publication as our KEY for
%% each reference.

\section{Introduction}

Studies of star formation are largely based on the observation of
molecular cloud cores.  We can estimate the density and temperature
distribution of such cores from dust continuum observations at
millimeter/submillimeter and infrared wavelengths.  Continuum
observations with SCUBA and ISO have especially improved our knowledge
of prestellar cores, which are cold dense cores without IRAS sources.
%which are objects where collapse has just begun and temperatures are
%still low.
Here the density
distribution has been found to be rather flat in the inner regions and
roughly consistent with a power law of the form $r^{-2}$ at larger
radii \citep{wt1994,awb2000,bac2000,shir2000,ev2001}.  As well as being an
independent probe of the density and temperature distribution,
molecular line emission can be a probe of the physical dynamics and
chemical kinetics in prestellar cores.  The interpretation of line
observations is more complicated than that of the continuum, however,
because molecular abundances may vary within a core and also with
time.  Recent observations have illustrated the different dependences
of molecular emission on position in a variety of prestellar cores.
For example, the ion N$_2$H$^+$ is relatively compact and shows a
well-defined peak while CO, CS, and CCS are more diffuse and sometimes
show a central hole.  Such emission patterns indicate that the latter
species are depleted at the center while N$_2$H$^+$ is not
    \citep{klv1996,wlv1998,cas1999,oha1999,lee2001}.
       Moreover, strong deuterium fractionation, an indicator of
the depletion of heavy molecules, has also been detected
\citep{hir1998,cas1999}.
An understanding of the chemistry occurring in prestellar cores would
appear to be essential to develop a detailed knowledge of the
conditions in the early stages of star formation.

Theoretical studies of the chemistry in prestellar cores have been
performed by several groups.  \cite{bl1997} calculated molecular
evolution in a fluid element in which the density increases as a
function of time.  These authors included gas-phase reactions and
adsorption and desorption of molecules on grain surfaces.  In their
model, species such as CO are depleted from the gas due to adsorption
onto grains, but N$_2$H$^+$ survives because of the lower binding
energy of its parent species N$_2$ on grain surfaces compared with
other heavy molecules.  \citet{aik2001} (hereafter Paper I)
utilized a similar chemical
model but with multiple fluid elements collapsing according to the
semi-analytical Larson-Penston solution \citep{lar1969,pen1969} in
order to obtain
the spatial distributions of molecules as a function of time.  The
column densities of CO, CS, and CCS were found to have a central hole,
while N$_2$H$^+$ was found to be centrally peaked.  Paper I
also investigated molecular column densities in more slowly collapsing
cores than the Larson-Penston flow to mimic the effect of magnetic
support or rotation.  It was found that the CCS abundance is
significantly lower in these cores, and that a Larson-Penston core
best reproduces the observed molecular abundances in L1544.  Since
molecular abundances are determined by a balance between chemical and
collapse time scales, the chemistry represents a probe of the collapse
time scale and thus of the physical mechanism of star formation.

\cite{li2002} went one step further, directly coupling a chemical
model of prestellar cores with a magneto-hydrodynamic simulation of
the dynamics.  The results of \citet{li2002} show that a core model
with magnetic support reproduces the observed molecular column
densities in L1544 better than a non-magnetized model.
The reasons for the different conclusions --- L1544 is best reproduced by
the Larson-Penston model in our Paper I but by the magnetized model
in \citet{li2002} --- are, however, not fully known.
The analysis of \citet{li2002} actually has two
advantages: inclusion of grain-surface reactions and direct coupling
between the chemistry and  magneto-hydrodynamic simulation via
ambipolar diffusion.
But, in order to include these two processes, \citet{li2002} were
forced to reduce the chemical network to save computational time.
%Moreover, these authors did not vary such important parameters as the
%sticking probability of gas-phase species on grain surfaces and
%the initial, pre-collapse density of the core.

This paper is an update of the earlier analysis of Paper I. Here we
investigate molecular distributions in a core collapsing according to
the Larson-Penston solution and its analogues with delayed collapse
using the new standard model (NSM) \citep{th1998} with deuterium
fractionation, and grain-surface reactions that are treated by the
modified rate approach \citep{sch2001}.  With the semi-analytical
Larson-Penston solution as a physical model, we can easily
incorporate the extensive chemical network needed.  There are three
goals of this paper.  Firstly, we hope to clarify what is the main
cause of the different conclusions between our previous work and
\citet{li2002}.  If the main cause is the inclusion of grain-surface
reactions, our present results will become similar to those of
\citet{li2002}.  Secondly, we will describe the detailed distribution
of gaseous and adsorbed molecules in assorted evolutionary stages of
prestellar cores in order to determine their value as probes of core
evolution.  Deuterated species are of special interest because the
D/H ratio has been used as a probe of the degree of ionization
\citep{cas1998} and the depletion of heavy gas-phase molecules
\citep{bm1989,rm2000}.  The distribution of adsorbed species in
prestellar cores has not been observed, but it provides an important
initial condition for warm protostellar cores, in which icy mantles
will be desorbed to start a second cycle of gas-phase reactions.
Thirdly, we will compare our results with observations of several
prestellar cores including L1544 and less dense prestellar cores such as
L1521E, which are often described as `starless cores' rather than 'prestellar
cores'.
Considering the uncertainties
both in observational data and chemical predictions, which will be
illustrated below, we need to compare observational data and
theoretical models for as many sources as feasible.
In addition, comparison of various prestellar cores with model
results should serve to tell us about the formation and 
evolution of cores.

The remainder of the paper is organized as follows.  The description
of our dynamical and chemical models is given in \S 2.  In \S 3, we
describe the evolution and distribution of molecular abundances in
collapsing cores.  We  compare our results with those of our
previous paper, in which grain-surface reactions were not included.
The dependence of our results on
collapse time scale, sticking probability, initial gas density, and
gas-phase chemical network is also investigated.  In \S 4, our model
results are compared with observational data, as well as with the results
of \citet{li2002}, while a summary is contained in \S 5.

\section{Model}

The distributions of density and velocity of the core, as obtained
from the Larson-Penston model, are shown in Figure 1 of Paper I at
several evolutionary stages.  Because the Larson-Penston solution is
self-similar, the ``shape'' of the density distribution is always the
same: it is almost flat at smaller radii, and follows a power-law $n
\propto r^{-2}$ at larger radii.  The initial central density of the
core, which is nearly homogeneous at radii through 10000 AU, is
assumed to be $n_{\rm H}=2\times 10^4$ cm$^{-3}$ (i.e.,
$n$(H$_2$)$=1\times 10^4$ cm$^{-3}$).
The core initially extends to the radius
of $3.2 \times 10^4$ AU, at which the density is
$n_{\rm H}=5.3\times 10^3$ cm$^{-3}$.
The subsequent collapse of
the core occurs in almost a free-fall time scale, with the central
density increasing to $n_{\rm H}=3\times 10^7$ cm$^{-3}$
and the outermost radius decreasing to $1.36\times 10^4$ AU
in $2.00\times 10^5$ yr. During the evolution, the core is isothermal with
$T=10$ K.
The Larson-Penston flow is an asymptotic
solution of a non-magnetized collapsing core; hydrodynamic simulations
and normal mode analysis
show that the flow asymptotically converges to the Larson-Penston flow
\citep{fos1993,hn1997}.  Although it is a good
approximation of collapsing dense cores especially in their inner
regions, the Larson-Penston flow has a problem at the outermost radii,
where the infall velocity is much higher than the observed value,
which is $\sim 0.1$ km s$^{-1}$.  One of the main causes for the discrepancy
is the outer boundary condition of the core.  In real objects, the infall
velocity of the outermost region should be zero, which is not
considered in this asymptotic solution.  Hence we cannot discuss
infall velocities and line profiles in our model.  Hydrodynamic
simulations with realistic (static) outer boundary conditions show,
however, that the spatial density distribution and temporal variation
of the central density of the Larson-Penston core are good
approximations to those in non-magnetized non-rotating cores
\citep{mmi1998}.

If a core is supported by rotation, magnetic fields, or turbulence,
collapse will be delayed, or the core will contract slowly.
Since molecular abundances are determined by a balance
between chemical and collapse time scales, the abundances in such a core
are different from those in the Larson-Penston solution.
In order to investigate the effect on molecular abundances,
we adopt analogues of the Larson-Penston core with delayed
collapse in which the infall velocity
is artificially decreased by a constant factor $f$. In comparison with
observation, we consider the fact that the factor $f$ need not be constant but
can vary with time in reality. In magnetized cores, the collapse is slower in
earlier, low-density, stages, while in rotating cores the centrifugal force
increases as the collapse proceeds.

Our chemical reaction network includes gas-phase reactions, gas-dust
interactions (i.e., adsorption and desorption), and diffusive
grain-surface reactions.  These surface reactions occur on grains of
radius $1.0 \times 10^{-5}$ cm that possess 10$^{6}$ binding sites.
All neutral species are assumed to be adsorbed when they hit the grain
surface with a sticking probability $S$ of unity, unless stated
otherwise.  Species on grain surfaces return to the gas phase via
thermal desorption and impulsive heating of the grains by cosmic-rays
\citep{ljo1985,hh1993}.  For the adsorption energies, we adopt those
of Model B in Paper I; these are basically the values calculated by
\citet{hh1993} but replaced, if available, with experimental data for
pure ices.
A set of grain-surface reactions was
taken from \citet{rh2000}.  To determine reaction rate coefficients,
those authors generalized the recent analysis of \citet{kat1999}, in
which a slow diffusion rate for H atoms on olivine and amorphous
carbon was deduced based on experimental measurements of HD formation.
Since we prefer to consider ice surfaces because these better represent
the mantles for much of the collapse, we calculate reaction rate
coefficients based on the older approach of \citet{hh1993}, in which
tunneling is included and activation energies against diffusion are
obtained as a constant fraction of adsorption energies.  Each surface
rate coefficient is proportional to the sum of the rates for the
two reactants to diffuse over the entire grain \citep{hh1993}.
These rates are then
altered if necessary according to the revised ``modified'' rate
approach of \citet{sch2001}, which can be described as follows.
Accretion, evaporation, and diffusion rates are calculated for
each reactant.  If the diffusion rate is the smallest among the three
values, the reaction rate coefficient is unaltered; otherwise the
diffusion rate in the rate coefficient is replaced by the larger
of the other two values.  The dominant alteration is to slow down the
reaction rates of atomic hydrogen and deuterium.  The modification
recently proposed for reactions with activation energy \citep{cas2002}
was not incorporated. The modified rate approach is a
simple manner of accounting for the discrete aspects of grain
chemistry, which are accounted for in greater detail by stochastic
methods \citep{ssh2002}.  These latter methods are not yet available
for large reaction networks.

The gas-grain reaction network has been extended to include singly
deuterated species.  As discussed in more detail in \S 3.7,
deuterium fractionation occurs in both the gas and surface phases.
Deuterium exchange reactions in the gas phase
with activation energies are adopted from \citet{mbh1989}.  We also
tried the newly measured and inferred rate coefficients for selected
fractionation reactions of \citet{ghr2002}.  The adsorption
energy and activation barrier against diffusion for surface D atoms are
assumed to be slightly higher than the values for H atoms
\citep{cas2002}.  In total our chemical network consists of 878
species and 11774 reactions.  In order to check the effect of
uncertainties in the network, we have also studied the core chemistry
with the UMIST gas-phase network \citep{ltmm2000} (\S 3.6), which we
have extended to include grain-surface reactions and deuterium species
in a similar way as described above.  In this case, 733 species and
11231 reactions have been considered.

As in Paper I, we have utilized the so-called ``low-metal'' values for
initial gas-phase elemental abundances (see Table 1 of Paper I) and
have adopted the ``standard'' value of $\zeta = 1.3 \times 10^{-17}$
s$^{-1}$ for the ionization rate by cosmic rays.
Considering that the  core is embedded in a molecular cloud, $A_{\rm v}=3$ mag
is assumed at the outer boundary of the core, so that  photodissociation
does not much affect our results.
All heavy elements are assumed to be
initially in atomic form, with carbon ionized
and oxygen neutral.  The initial form of hydrogen is molecular, and
the deuterium is assumed to be in the form of HD, at a ratio of $3.0
\times 10^{-5}$ with respect to H$_{2}$.

\section{Results}

\subsection{Molecular Evolution in an Infalling Fluid Element}

\begin{figure}
\epsscale{0.8}
\plotone{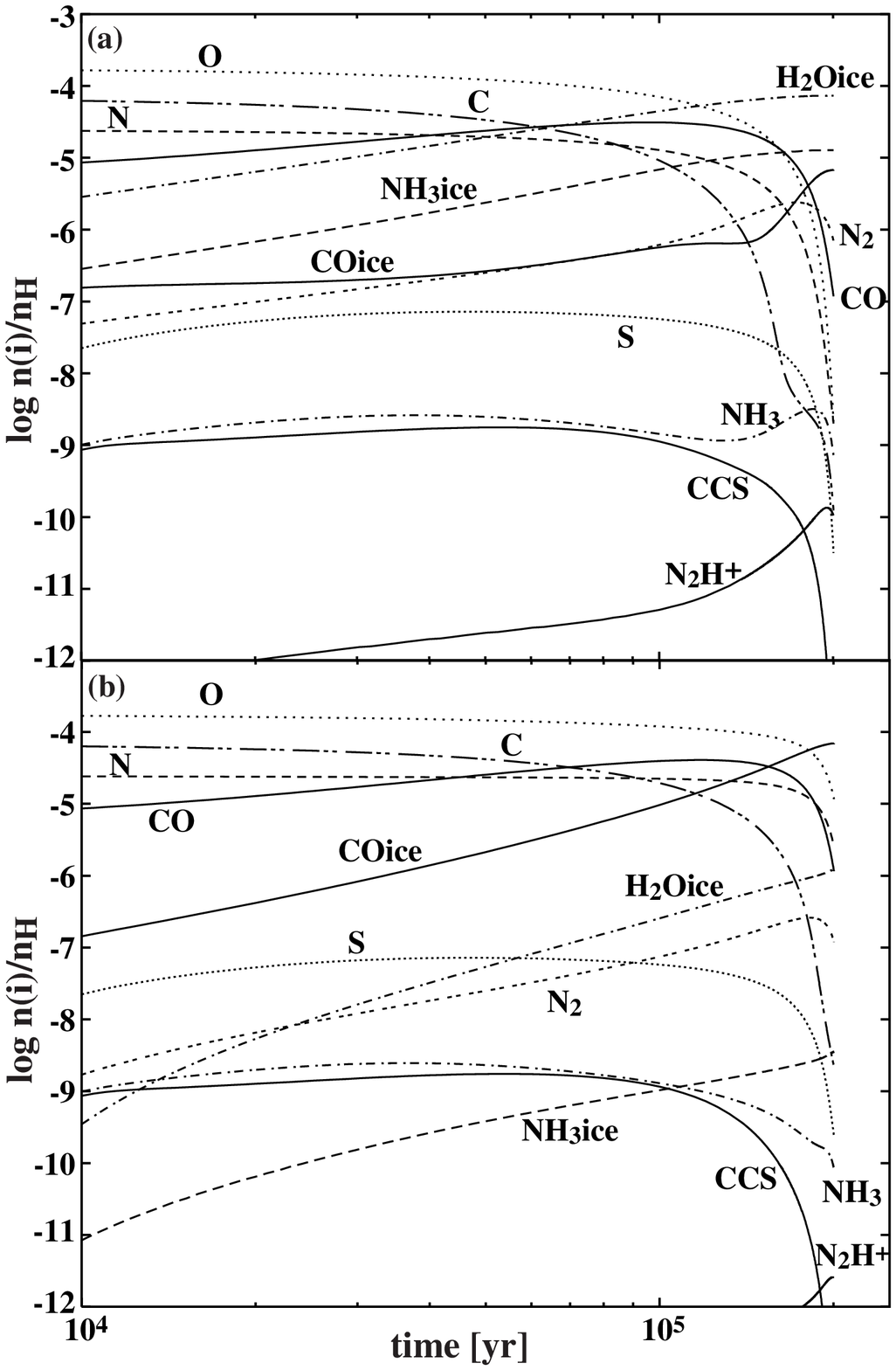}
\caption{Temporal variation of molecular abundances relative to hydrogen nuclei
in a fluid element that migrates from $8.2
\times 10^3$ to $1.0\times 10^3$ AU in $2.00\times 10^5$ yr.
Only the later stages of the evolution are shown. The top panel (a)
depicts the results of our current paper with grain-surface
reactions, while the bottom panel (b) shows results obtained without
grain-surface reactions, as in Paper I.
\label{fig:evol}}
\end{figure}

Figure \ref{fig:evol} (a) shows the molecular evolution in terms of
fractional abundances ($n(i)/n_{{\rm H}}$) that occurs in a fluid
element that migrates from a radius of $8.2 \times 10^3$ to one of
$1.0\times 10^3$ AU in $2.00\times 10^5$ yr, while the gas density
$n_{\rm H}$ increases from $1.7\times 10^4$ to $5.8\times 10^6$
cm$^{-3}$.  Species on the surfaces of dust grains are designated as
ices.  Only the later stages of collapse are depicted because both the
density variation and molecular evolution are rapid after $10^5$ yr.
At this latest stage, the abundances of CCS and CO decrease with time
while that of N$_2$H$^+$ increases.  The reasons for these trends are
partially the same as described in previous papers and partially new;
CCS, a so-called ``early-time'' species is depleted via gas-phase
reactions and, following adsorption, reactions on grain surfaces, while CO
is adsorbed onto grains, where it reacts to produce
more complex organic species.
The ion N$_2$H$^+$ increases in abundance for at least two
reasons: (i) the binding
energy of its parent molecule N$_2$ on grain surfaces is low, which
allows its replenishment into the gas phase, and (ii) its rate of
depletion is slowed when reactants such as CO stick to the grains
(Bergin \& Langer 1997; Paper I).  Meanwhile, the abundances of the major ices
-- water, ammonia, and CO -- increase with time.  Methane (CH$_{4}$)
and CO$_2$ ices are formed as well, and reach abundances as high as
$n(i)/n_{\rm H}\sim 10^{-5}$, although they are not shown in the
figure.

Figure \ref{fig:evol} (b) shows the molecular evolution occurring in
the same fluid element as in (a), but with grain-surface reactions
inactivated except for H$_2$ formation and ion-electron recombination.
The results presented are essentially those of Paper I. Several
major differences in molecular evolution are immediately apparent
between the result presented in Figure \ref{fig:evol} (b) and those of
our standard model (a).  First, the ices that are formed mainly via
surface reactions have greatly reduced abundances; among these are
water, ammonia, carbon dioxide and methane.
Secondly, carbon monoxide ice, formed
mainly via gas-phase reactions followed by accretion, is more
abundant in Figure \ref{fig:evol} (b),
%partially because the slow surface
%reaction with atomic H is turned off.
because the transformation to other species via surface reactions is
turned off.
Thirdly, gas-phase species can
also have altered abundances.  The nitrogen molecule has a lowered
abundance in Figure \ref{fig:evol} (b) because its formation on grain
surfaces followed by desorption into the gas is significant in the
normal model (a).  The lowered abundance of molecular nitrogen leads to
the low abundance of N$_{2}$H$^{+}$.
The depletion of CCS is slightly accelerated because more
oxygen is in atomic form rather than in the form of water ice; atomic
oxygen aids in the transformation of ``early-time'' species such as
CCS to CO.

\subsection{Distribution of Molecular Abundances}

\begin{figure}
\epsscale{1.05}
\plotone{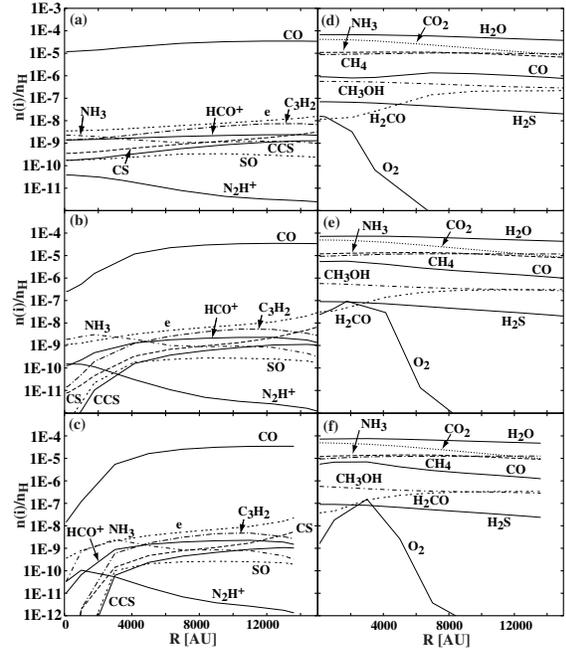}
\caption{Distributions ($n(i)/n_{\rm H}$) of molecules
at $t= 1.52 \times 10^5$ yr,
$1.89\times 10^5$ yr, and $2.00\times 10^5$ yr (from top to bottom),
when the central density of the core is $3\times 10^5$ cm$^{-3}$,
$3\times 10^6$ cm$^{-3}$, and $3\times 10^7$ cm$^{-3}$,
respectively.  The left panels show distributions of gaseous molecules,
while the right panels show those of adsorbed species (``ices'') on
grain surfaces.  \label{fig:dist}}
\end{figure}

The radial distributions of molecular fractional abundances within the
collapsing core at assorted times are obtained by solving for the
molecular evolution in a similar manner to that described in the
previous section, but for many fluid elements in the flow.  The left
panels in Figure \ref{fig:dist} show distributions of gaseous species
at $t= 1.52 \times 10^5$ yr, $1.89\times 10^5$ yr, and $2.00\times
10^5$ yr (from top to bottom).  At these times, the central density of
the core is $3\times 10^5$ cm$^{-3}$, $3\times 10^6$ cm$^{-3}$, and
$3\times 10^7$ cm$^{-3}$, respectively.  The distributions of gaseous
species are qualitatively similar to those of Paper I: they are
relatively flat at the early stage, labeled (a), but the radial
dependence becomes more apparent at later stages, labeled (b) and
(c).  When the central density is $3\times 10^6$ cm$^{-3}$, NH$_3$ and
N$_2$H$^+$ become more abundant in the inner regions, while the other
gaseous species are depleted by orders of magnitude at the center.
When the central density reaches $3\times 10^7$ cm$^{-3}$, even
ammonia and N$_2$H$^+$ show a central depletion.

The right panels in Figure \ref{fig:dist} show distributions of
adsorbed species on the grains.  Variations among the three
evolutionary stages are less pronounced compared with the gaseous
species.  Water ice is always the dominant component in the grain
mantle, while the NH$_3$ and CH$_4$ ice abundances are 10-30\%
relative to water.  These three species are formed predominantly by
surface reactions in which atomic hydrogen converts adsorbed oxygen,
nitrogen, and carbon atoms into the saturated species.  Adsorbed CO,
formed mainly in the gas phase, is transformed to CO$_2$, H$_2$CO, and
CH$_3$OH.

\begin{figure}
\epsscale{0.8}
\plotone{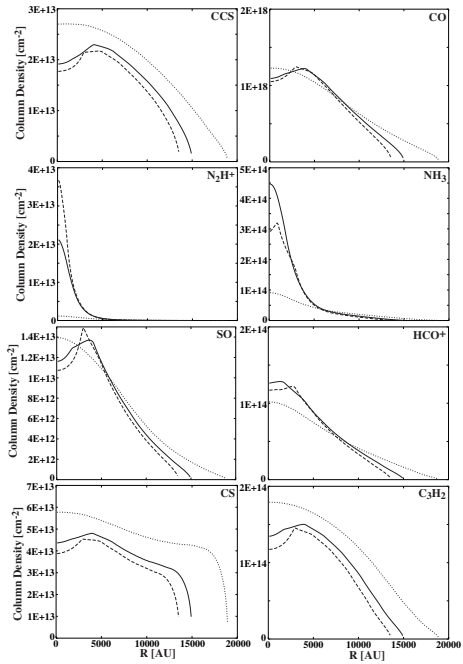}
\caption{Molecular column densities as a function of impact parameter
(radial distance) from
the core center when the central density of the core is $3\times 10^5$
cm$^{-3}$ (dotted lines), $3\times 10^6$ cm$^{-3}$ (solid lines), and
$3\times 10^7$ cm$^{-3}$ (dashed lines). \label{fig:column_time}}
\end{figure}

Assuming that the core is spherical, we integrate the molecular
abundances along the line of sight to obtain molecular column
densities as a function of impact parameter (radial distance) from the
core center at the three representative evolutionary stages.  The
results are shown in Figure \ref{fig:column_time}.  While the central
density increases from $3\times 10^5$ cm$^{-3}$ to $3\times 10^7$
cm$^{-3}$, the total (hydrogen) column density is enhanced by about an
order of magnitude.  The central molecular column densities shown
in Figure \ref{fig:column_time}, however, show only small changes, except for
N$_2$H$^+$ and NH$_3$.
%{\sc Yuri: do the small changes refer to all impact
%parameters or to the central column densities?}
  The column densities of most neutral species,
such as CO, CS, SO, CCS, and C$_3$H$_2$, are initially centrally
peaked, but develop a central hole at later stages, while N$_2$H$^+$
is centrally peaked even in the latest stage, and ammonia shows a
very tiny central hole at the latest stage.  These qualitative
features are the same as those reported in Paper I. But the column
densities of N$_2$H$^+$, NH$_3$, and C$_3$H$_2$ show significant
enhancements from those in Paper I, which were computed without
grain-surface reactions.  If we consider the time when the central
density of the core is $3\times 10^6$ cm$^{-3}$, the peak column
densities of N$_2$H$^+$, NH$_3$, and C$_3$H$_2$ in Paper I are
$2.5\times 10^{11}$ cm$^{-2}$, $3\times 10^{13}$ cm$^{-2}$, and
$2\times 10^{13}$ cm$^{-2}$, respectively, about an order of magnitude or
more lower than reported here.  The increase in the ammonia column
density stems from the grain-surface formation of N$_{2}$ which,
unlike ammonia ice, is followed by desorption and a sequence of
gas-phase reactions leading to gaseous NH$_{3}$.  The analogous
increase in N$_2$H$^+$ also follows from the grain-surface formation
and subsequent desorption of N$_2$.  Unlike these two cases, the
enhancement in the column density of C$_3$H$_2$, an ``early-time''
species, is the result of a depletion of O atoms (\S 3.1).

Now that we have summarized the results of our standard model, it is
useful to see how these results depend on some important parameters,
such as the time scale for collapse, the sticking probability on
grains, the initial conditions, and the reaction network.

\subsection{Collapse Time Scale}

\begin{figure}
\plotone{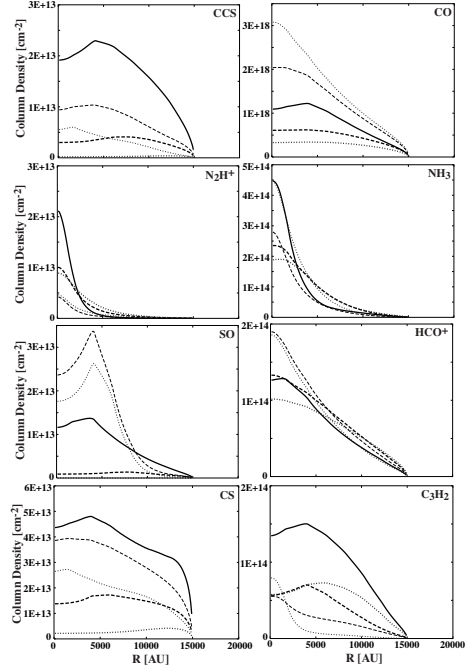}
\caption{ Molecular column densities as a function of impact
parameter when the central density of the core is
$3\times 10^6$ cm$^{-3}$. The solid lines represent the Larson-Penston core,
while the dashed and dotted lines represent cores collapsing more
slowly by factors $f$ of 3 and 10, respectively. The sticking probability
is 1.0 for all thick lines, and $1/f$ for thin lines. \label{fig:cf_slow}}
\end{figure}

The thick solid lines in Figure \ref{fig:cf_slow} show molecular
column densities as a function of impact parameter for the case of the
Larson-Penston core ($f=1$), while cores which collapse more slowly
than the Larson-Penston model by a factor $f$ of 3 and 10 are depicted
by thick dashed and dotted lines, respectively.  The central densities
of the cores are all at a density of $3\times 10^6$ cm$^{-3}$, so they
would appear the same if observed via the dust continuum.  As in Paper
I, the molecular column densities are generally smaller in the more
slowly collapsing cores because of heavier depletion (e.g., CCS, CO,
SO, and CS).  The neutral molecule NH$_3$ and the
molecular ion N$_2$H$^+$ show a more complex pattern: here the more
slowly collapsing core leads to larger column densities at larger
radial distances but smaller column densities near the center of the
core.
At outer radii, the parent molecule N$_2$ is more abundant
in the slow collapse models
because of its slow formation rate, while at inner radii, on the other hand,
more nitrogen is depleted onto grains as NH$_3$ ice in the slow collapse
models.

In general, the sensitivity of the column densities to the collapse
time scale is smaller than that obtained in Paper I. For example, the
central column densities of CCS and C$_3$H$_2$ in the $f=3$ model are
lower than those in the $f=1$ model by more than an order of magnitude
in Paper I, whereas here they are lower by factors of 6 and 2,
respectively.  What causes this change?  The abundances of the
early-time species, which are typically unsaturated (H-poor) organic
molecules such as CCS and C$_{3}$H$_{2}$, are sensitively dependent on
the C/O elemental ratio in the gas phase \citep{prat1997,th1998}
according to gas-phase models of cold clouds because this elemental
ratio determines how much atomic oxygen is available in the gas.  In
the current models, early-time species are relatively abundant even in
the slowly collapsing cores because of the depletion of O atoms via
water formation on grain surfaces.  The dependence of the NH$_3$ and
N$_2$H$^+$ column densities on the collapse time scale is also smaller
than in Paper I, where the central densities of NH$_3$ and N$_2$H$^+$ in
the $f=10$ model are larger than those in the $f=1$ model by factors
of 5 and 10, respectively, because of their slow formation in the gas
phase.  Here, the major synthetic pathways for both species lie
through N$_{2}$ that is desorbed from grain surfaces.  The increase
in time available due to slow collapse has less of an effect.

\subsection{Sticking Probability}

The thin lines in Figure \ref{fig:cf_slow} show molecular column densities
vs. radial distance for slowly collapsing cores with low sticking
probabilities $S$.  Specifically, for an $f$ of 3 (dashed lines), $S$ is
0.33, and for an $f$ of 10 (dotted lines), $S$ is 0.1.  The
lowered sticking probabilities are chosen to compensate for the
increased time spent in collapse.  The compensation is of course only
a partial one, because rate equations are non-linear, including many
reactions other than adsorption.

The dependence on the sticking probability varies among species.  The
column density of CO is higher in models with a lower sticking
probability.  Grain-surface reactions transform CO to CO$_2$, H$_2$CO,
and CH$_3$OH, which mostly remain on the grain because of higher
adsorption energies.  In the models with lower sticking probability,
such processes are less efficient.
The column density of SO is higher in models with a lower sticking
probability as well, because depletion of SO itself is slow enough to
allow the late peak to grow after the depletion of carbon atoms,
which are the dominant reactant during the early-time chemistry.
Although the gas-phase abundances of CCS, CS, and C$_3$H$_2$ are
highest in the inner regions for the standard Larson-Penston case, for
models with $f > 1$ the abundances are higher in models with a lower
sticking probability and thus with a smaller rate of depletion.  At
larger radii, on the other hand,
%they are less abundant in the models
C$_3$H$_2$ is less abundant in the models
with a lower sticking probability, because of the higher abundance of
destructive O atoms in the gas phase.
Gas-phase N$_2$ and NH$_3$ tend
to increase in abundance at small radii with lower sticking rates,
because the lessened rate of depletion via grain adsorption
is more important than the lessened rate of formation, while
N$_2$H$^+$ is less abundant because of the higher abundance of CO,
reactions with which deplete the ion.

\subsection{Initial Conditions}

\begin{figure}
\plotone{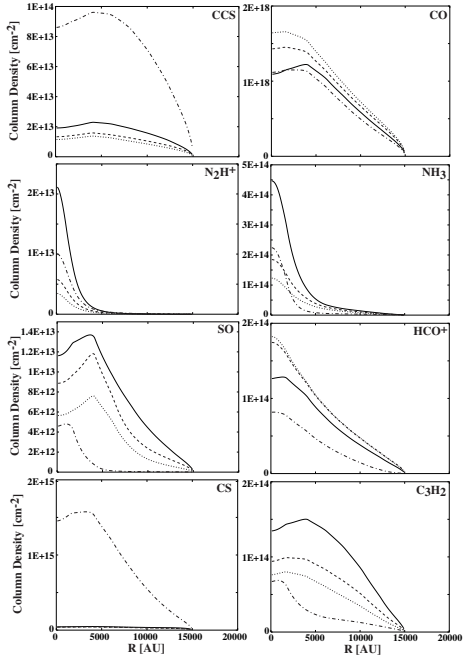}
\caption{Molecular column densities in the Larson-Penston core when the
central density is $3\times 10^6$ cm$^{-3}$. The initial central density of
the core (cm$^{-3}$) is $2\times 10^4$ for the solid lines, $2\times 10^3$ for
dashed lines, and $2\times 10^2$ for dotted lines.
The dashed-dotted lines show the molecular column densities obtained using
the UMIST-based chemical reaction network.\label{fig:tau_UMIST}}
\end{figure}

So far we have assumed that the initial central density of the core,
$n_{\rm H}(t=0, R=0)$, is $2\times 10^4$ cm$^{-3}$, which is often
smaller than observed in quiescent dense cores but is often found in
molecular clouds (e.g. Onishi et al. 1996). Since the initial condition
of the core is uncertain, it is interesting to
determine the dependence of the molecular column densities on the
initial gas density. Figure \ref{fig:tau_UMIST} shows molecular column
densities in cores with initial central densities of $2\times
10^4$ (solid lines), $2\times 10^3$ (dashed lines), and $2\times
10^2$ cm$^{-3}$ (dotted lines).  In all three models, the initial gas
consists of our standard conditions: atoms and ions except for
hydrogen, which is in molecular form.  The central density in all
models is $3\times 10^6$ cm$^{-3}$, which is reached after $1.89\times
10^5$ yr, $6.33\times 10^5$ yr, and $2.04\times 10^6$ yr,
respectively.  It should be noted that the latter two models are in a
low density state for a significant fraction of the time; their
central densities reach $2\times 10^4$ cm$^{-3}$ at $t=4.44\times 10^5$ yr
and $1.85\times 10^6$ yr,
after which the evolution of the density distribution is the same as
that of the first model with the time $t$ of calculation shifted
accordingly.

It can be seen in Figure \ref{fig:tau_UMIST} that the dependence on
the initial conditions is a factor of $\sim 2$ for most of
the species, which suggests that the early evolutionary stages with
low densities do not much affect the molecular abundances in the dense
core.  Exceptions are N$_2$H$^+$ and NH$_3$, which are less abundant
in the models with a low initial density.
In these models, the
higher abundance of H atoms enhances the surface formation of NH$_3$
at the expense of N$_{2}$, which is the major precursor of gaseous
ammonia as well as N$_{2}$H$^{+}$, since surface ammonia does not
desorb efficiently in our model.  In reality, the
dependence of the N$_2$H$^+$ and NH$_3$ abundances on the initial
density could be smaller.  In clouds with low density ($\sim 10^2$
cm$^{-3}$) the temperature would be higher than assumed here (10
K), and hence efficient desorption of H atoms would suppress the
surface formation of NH$_3$.

\subsection{Uncertainties in Reaction Networks}

Molecular evolution proceeds via a chemical reaction network that
consists of a large number of reactions both in the gas phase and on
the grain surfaces.  The gas-phase chemistry has been studied by more
investigators over a longer period of time, and is generally better
understood than the grain-surface chemistry.  Gas-phase models have
succeeded in reproducing or predicting the abundances of most (80 \%)
of the observed species in quiescent sources, in understanding
four-order-of-magnitude deuterium enrichments, and in explaining the
existence of peculiar species such as carbon-chain molecules
\citep{th1998,mbh1989,rm2000}.  Yet, in spite of these successes and
significant efforts in refining the networks, uncertainties in the
chemistry still remain.  These uncertainties arise from a variety of
causes.  First, most reactions in the models have not been studied in
the laboratory and their rates are estimated based on analogous
better-studied systems.  These estimates are on somewhat firmer ground
for ion-molecule reactions than they are for neutral-neutral
reactions, since it is always difficult to determine whether the
latter possess significant activation energies.  The extent of the
role of neutral-neutral reactions remains a significant problem
\citep{th1998}.
Secondly, most applications of the chemical models are at low
temperatures (e.g., 10 K) whereas most laboratory experiments are at
higher temperatures.  The temperature dependence can often be obtained
from theoretical considerations or comparison with other systems measured
over a large temperature range, but not always with certainty.  A
simple example concerns the rate coefficients for ion-polar neutral
reactions, for which some authors include temperature-independent rate
coefficients while others, extrapolating from theory and a few
measured results, include an inverse temperature dependence.  The
strong but uncertain temperature dependence of radiative association
reactions also leads to significant uncertainties.  Thirdly, even subtle
effects such as the role of fine structure can make a conversion from
a laboratory rate to an interstellar rate rather uncertain.

Currently there are two partially-independent extensive gas-phase
chemical reaction networks: the UMIST data base \citep{ltmm2000} and
the New Standard Model (NSM) \citep{th1998}.  Since the authors of
these networks have different policies and standards in compiling the
data, the differences in the two models can give us a sense of how
uncertainties in chemistry affect the results.  But this is only a
sense, because the absolute uncertainties in reaction rate
coefficients probably exceed in many instances the differences between
the two networks.  \citet{ma2002}  has used a
Monte Carlo procedure to show how the estimated uncertainties in the
latest rates adopted by the UMIST compilers lead to uncertainties in
the predicted abundances, an effect which becomes significant for
large molecules.

We have adopted the NSM for the gas-phase reactions in our calculation.
Here we replace it with the UMIST data base, and check how significantly our
results are affected by uncertain rate coefficients.
The dash-dotted lines in Figure \ref{fig:tau_UMIST} show the molecular column
densities obtained with the UMIST data base, which we extended to include
the same grain-surface reactions and gas-grain interactions as in our
NSM-based model. The physical model is the Larson-Penston flow with an initial
central density of $2\times 10^4$ cm$^{-3}$. Comparison with
results obtained with the NSM (solid
lines in Figure \ref{fig:tau_UMIST}) suggests that the uncertainties in the
gas-phase network cause errors of a factor of $\lesssim 2$ in the column
densities of most molecular species. Exceptions are sulphur-bearing molecules;
the CCS and CS column densities are higher with the UMIST data base than with
the NSM by factors of 4 and 30, respectively. These differences are not
surprising because it has been known for some time that the differences
among S-bearing species exist even in the purely gas-phase pseudo-time
dependent (i.e. constant density) models.  In an effort to locate the
specific source of the discrepancies,
we checked the main formation and destruction paths for each molecular species
and found several important reactions which have different rate coefficients
in the two networks.

One of the critical reactions is
O + HCS$^+$, since this reaction depletes protonated CS and interferes
with the production of CS via dissociative recombination with electrons.
The rate coefficient for this reaction is $5.0\times 10^{-12}$ cm$^3$ s$^{-1}$
in the UMIST data base
and $5.0\times 10^{-10}$ cm$^3$ s$^{-1}$ in the NSM.  It is of interest
to illuminate how this difference came about.  Since the UMIST
website (http://www.rate99.co.uk/)  shows how the rate
coefficients for individual reactions are obtained, one can see that
the rate coefficient for O + HCS$^{+}$ stems from an early paper by
\citet{mh1990} in which it is actually written that ``both rate
coefficients and products are uncertain.''  The NSM value is
presumably based on other studies involving atomic oxygen, which tend
to show rate coefficients up to one order-of-magnitude below the
Langevin rate.  Another critical reaction is that between C and SO.
The UMIST value stems from a simple theoretical model used in the
shock study of \citet{lg1988} while the origin of the much lower NSM
value is uncertain.
     From these examples, one can see that different choices can be made
when modelers are faced with less than complete information.
Regarding the very uncertain rates of neutral-neutral reactions, one
of us (E.H.) has recently started to collaborate with a European
network of chemical kineticists to obtain expert opinion on unstudied
systems.  The resulting rate coefficients for neutral-neutral
reactions will no longer contain what once was the standard
temperature dependence of (T/300)$^{0.5}$ derived from hard sphere
arguments.

\subsection{D/H ratio}

The elemental abundance of deuterium is $\sim 10^{-5}$ relative to
hydrogen in the local interstellar medium.  In molecular clouds,
deuterium is predominantly in the form of HD, but is transferred from
this reservoir to other molecular species by proton/deuteron exchange
reactions, the most important of which is H$_3^+$ + HD $\to$ H$_2$D$^+$ + H$_2$
\citep{mbh1989,rm2000}.  The exchange reactions are
exothermic by amounts of energy large enough that the backwards
endothermic reactions are inefficient in clouds with temperatures
significantly lower than the reaction energy expressed as a
temperature.  As a result, D/H ratios such as H$_2$D$^+$/H$_3^+$ are
much larger than the elemental D/H ratio, an effect known as
fractionation.  The deuterated isotopomers formed in exchange
reactions react with other species to transfer the deuterium further,
producing a significant deuterium fractionation for many minor species
in the model.  The ion H$_2$D$^+$
%and analogous ions (H$_{2}$DO$^{+}$, CH$_{4}$D$^{+}$) are
is depleted at low temperatures principally by
reactions with electrons and neutral species with high proton affinity
such as CO and H$_2$O. If these heavy species are depleted from the
gas onto the dust grains, then the D/H ratios in H$_3^+$ and in other
molecular species are enhanced if, as is the case here, the ionization
degree is low (i.e. $n$(e)$/n_{\rm H}\lesssim 3 \times 10^{-7}$).
These arguments suggest that D/H ratios of species
other than HD can be used as probes of cloud conditions such as
temperature, degree of ionization, and degree of molecular depletion.
It is therefore useful to present molecular D/H ratios for our models.

Fractionation can also occur via grain-surface chemistry.  In
particular, D atoms are formed in the gas phase via the dissociative
recombination of molecular ions such as H$_{2}$D$^{+}$ and DCO$^{+}$,
and then are adsorbed onto grain surfaces, where they can react in
much the same manner as atomic hydrogen.  For example, if a surface
CO molecule is subjected to reactions with atomic H and D, both the
normal species H$_{2}$CO and CH$_{3}$OH can be produced as well as
their singly and multiply deuterated isotopomers
\citep{ctr1997,cas2002}.

\begin{figure}
\epsscale{0.9}
\plotone{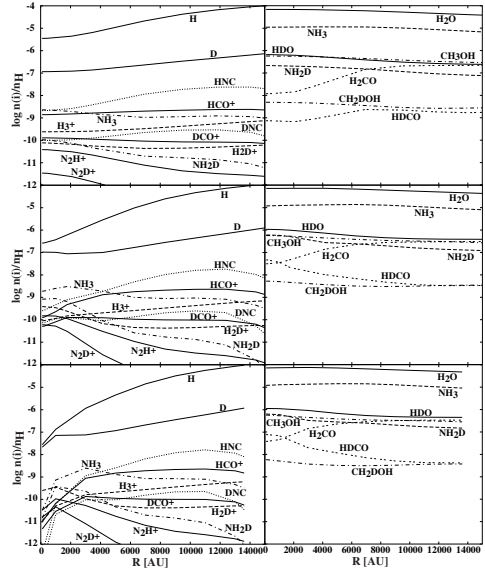}
\caption{Radial distributions of the abundances of deuterated and normal
species in the
Larson-Penston core when the central density of the core is $3\times
10^5$ cm$^{-3}$ (top), $3\times 10^6$ cm$^{-3}$ (middle), and $3\times
10^7$ cm$^{-3}$ (bottom). The left panels show the abundances of
gas-phase species, and the right panels show those of adsorbed
species on grain surfaces.\label{fig:dist_D}}
\end{figure}

Figure \ref{fig:dist_D} shows radial distributions ($n(i)/n_{\rm H}$)
of deuterated and normal species in the Larson-Penston core when the
central density of the core is $3\times 10^5$ cm$^{-3}$ (top),
$3\times 10^6$ cm$^{-3}$ (middle), and $3\times 10^7$ cm$^{-3}$
(bottom).  The left panels show gas-phase species, and the right
panels show abundances in the grain mantles.  As expected, the D/H
ratios in the gas-phase molecules are higher at smaller radii, and at
later evolutionary stages because of the stronger depletion of heavy
molecules and the lowered ionization degree. The ionization degree obtained
in the calculation can be approximately fitted by the simple function
$6\times 10^{-9} (10^5{\rm cm}^{-3}/n_{\rm H})^{0.5}$
%{\sc Yuri: don't you want the `cm-3' following nH?}
in regions with
relatively high density ($n_{\rm H} \gtrsim 10^5$ cm$^{-3}$) for the
Larson-Penston model, and varies by a factor of $\lesssim 2$ depending on the
parameter $f$ and $S$. This degree of ionization is close to
what one estimates if all ionization stems from the interaction of
cosmic rays and molecular hydrogen, and is also consistent with the value
($\sim 2\times 10^{-9}$) estimated by \citet{cas2002b} at the center of L1544
where $n_{\rm H}\sim 10^6$ cm$^{-3}$.
The resulting
deuterium fractionation can be so high that the deuterated isotopomers
can become more abundant than the normal isotopomers;
H$_2$D$^+$ is more abundant than H$_3^+$, and is the most abundant ion
in the innermost regions when the density is high. Given the high degree
of deuterium fractionation, further studies with multiply deuterated
isotopomers would be interesting in relation to the recent observation of these
species in protostars.

Among the surface species, the radial distributions show somewhat less of a
dependence on density and radius.  Still, there are exceptions,
particularly singly deuterated formaldehyde - HDCO - which, for the
higher density inner cores, actually becomes more abundant than the
normal isotopomer.  On the other hand, the more abundant of the two
singly deuterated isotopomers of methanol - CH$_{2}$DOH - only
achieves an abundance approximately 0.01 of normal methanol.

\begin{figure}
\epsscale{1.0}
\plotone{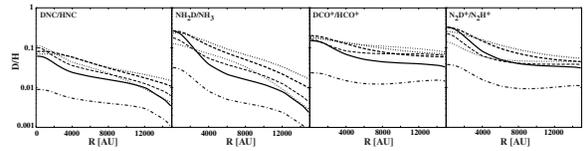}
\caption{Column density ratios of some deuterated isotopomers to
normal species as a function of impact parameter from the core center
when the central density is $3\times 10^6$ cm$^{-3}$.  Solid lines
represent the Larson-Penston core, while dashed and dotted lines
represent cores collapsing more slowly by factors $f$ of 3 and 10,
respectively.  The sticking probability is 1.0 for thick lines, and
$1/f$ for thin lines.  Dashed-dotted lines show the molecular column
densities in the Larson-Penston core, but with the ``New Rates'' for
deuterium fractionation reactions.\label{fig:col_ratio_D}}
\end{figure}

So far, we have discussed the radial distribution in terms of
fractional abundances and not the column
densities of deuterated and normal isotopomers as functions of impact
parameter.  Figure \ref{fig:col_ratio_D} shows column density ratios
for some deuterated species to normal species as a function of impact
parameter from the core center when the central density is $3\times
10^6$ cm$^{-3}$.  The Larson-Penston core is represented by solid
thick lines, while collapses slowed down by factors $f$ of 3 and 10
are depicted by thick dashed and dotted lines, respectively.  For
these latter two cases, models were also run with sticking
probabilities of $1/f$; the results are, as usual, shown as thin dashed and
dotted lines.  As can be seen, the D/H ratios are higher in models
with heavier depletion; i.e. those with slow collapse and a high
sticking probability.  The ratio, which is an average along the line
of sight, also varies with species; it is higher if the radial
distribution of the species is more centrally peaked because deuterium
fractionation is greater in the inner regions.  For example, the
molecular ion N$_2$H$^+$ and the ``late-time'' species NH$_{3}$ have
higher ratios than neutral ``early-time'' species such as HNC.

In the models presented above, the rate coefficients for the deuterium
fractionation exchange reactions are adopted from \citet{mbh1989}.
Recently, \citet{ghr2002} and \citet{gs2002} measured these
rate coefficients in an ion trap at 10 K, and found that the rate
coefficients of the forward reactions (e.g. H$_3^+$ + HD $\to$ H$_2$D$^+$
+ H$_2$) are smaller and those of backward reactions probably larger
than those utilized by \citet{mbh1989}.
\citet{rhm2002} showed that these revisions of the rate coefficients
significantly lower the D/H ratios in quiescent cores for both
gas-phase and accretion-containing models.  To determine the effect of
the revised rates for prestellar cores, we used theses rates
(summarized in Table 1 of \citet{rhm2002}) with a core undergoing
Larson-Penston collapse.  Some results are shown with dashed-dotted
lines in Figure \ref{fig:col_ratio_D}.  The D/H ratios are typically
lowered by an order of magnitude.

\section{Discussion}

\subsection{Comparison with L1544 and Previous Models}

The prestellar core L1544 is the most intensely studied object of this
class.  Its radius is about 15000 AU, and the central density
$n$(H$_2$) is estimated to be $1.5\times 10^6$ cm$^{-3}$ from dust
continuum measurements.  Various molecular lines have been
observed and gas-phase molecular column densities estimated.
Detailed mapping and interferometer
observations have revealed that CO and CCS are depleted at the core
center, while the N$_2$H$^+$ emission peaks at the center
\citep{cas1999,oha1999,taf1998,wil1999}.

Firstly, we compare molecular column densities obtained in our models with
those observed at L1544.
In Table \ref{L1544}, we list the estimated molecular column densities from
observational data which are obtained at the core center.
%peak position of dust continuum or N$_2$H$^+$ emission.
Table \ref{L1544} also lists molecular column
densities calculated in our models
with a variety of $f$ and $S$ values when the central density is
$n_{\rm H}=3\times 10^6$ cm$^{-3}$.  The tabulated results are
surface-averaged (i.e. weighted by 2$\pi R dR$) within a radius
of 1000-2000 AU depending on the actual beam size of each observation.
The bold numbers indicate
agreement with the observation to within a factor 3, while the
numbers in italics
indicate disagreement by a factor of $>5$.  The model with $f=1$
shows the best agreement, while the very slow collapse model ($f=10,
S=0.1$) shows the worst agreement.
%The spatial distributions of
%molecules --- the central depletion of CCS and CO, and the
%centrally-peaked distribution of N$_2$H$^+$ --- are well reproduced in
%the Larson-Penston core and only somewhat less so with other models
%(Figure \ref{fig:cf_slow}).

The factor $f$, the ratio of the actual collapse time scale to the
Larson-Penston collapse time scale, should be considered with caution.
Although we set $f$ to be constant for simplicity, in reality it
varies with time.  If the core is supported by magnetic fields, $f$ is
large in the early low-density stages, but decreases to unity when the
core becomes too dense to couple with magnetic fields.  On the other
hand, $f$ increases in the later stages of collapse via rotational
support, if the core has significant angular momentum.  Since chemical
reactions generally proceed faster at higher densities, the value of
the factor $f$ at higher densities is more important in determining
the molecular abundances in cores.  Therefore the fact that the worst
agreement is obtained with the highest value of $f$ indicates that the
collapse time scale {\it at a central density of $\sim 10^6$
cm$^{-3}$} should not be larger than the free-fall time scale by a
factor of 10.

\begin{figure}
\epsscale{0.9}
\plotone{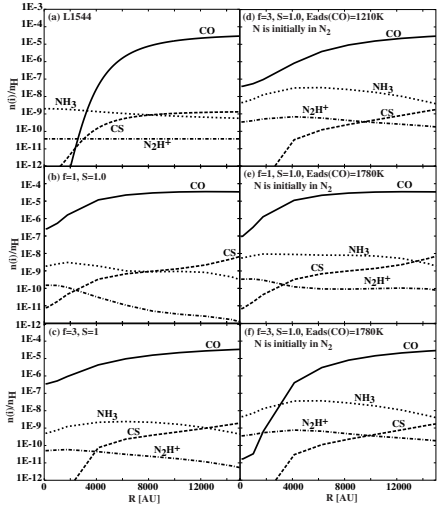}
\caption{Radial distribution of molecular abundances in L1544
estimated by \citet{taf2002} (a), and obtained in our models when the
central density of the core is $n_{\rm H}=3\times 10^6$ cm$^{-3}$ (b-f).
Part (b) shows results from our Larson-Penston core, while the slow collapse
model ($f=3$ and $S=1$) is shown in (c). Parts (d-f)
show results from revised models with nitrogen initially in its
molecular form and a higher adsorption energy for CO.
\label{fig:dist_L1544}}
\end{figure}

Secondly, we compare the radial distribution of fractional
molecular abundances in our model cores with those in L1544.
\citet{taf2002} utilized radiative transfer calculations to estimate
the distribution of fractional molecular abundances in
prestellar cores such as L1544 from observational data.  These
distributions, which represent a new development, are not to be
confused with column density measurements. Figure
\ref{fig:dist_L1544} (a), taken from formulae of \cite{taf2002},
shows their radial distribution of molecular abundances in L1544.  At
the core center, CO and CS are heavily depleted and NH$_3$ is slightly
enhanced, while the abundance of N$_2$H$^+$ is constant.
Figures \ref{fig:dist_L1544} (b) and (c) show our results for the L-P core
($f=1, S=1$) and the model with $f=3$ and $S=1$ when the central density of
the core is $3\times 10^6$ cm$^{-3}$. The calculated distributions are in
qualitative agreement with those determined in L1544, but not in quantitative
agreement, except for NH$_3$. In our models, the N$_2$H$^+$ 
fractional abundance is significantly
higher towards smaller radii, while it is inferred to be constant in L1544.
Also, the gradient of the CO abundance in the central region of L1544 
is steeper
than in our models.  Although the steep CO gradient in L1544 is partly caused
by the exponential function assumed in the analysis, the CO intensity profile
obtained from our L-P core is still more centrally peaked than in L1544
(Tafalla 2003, private communication).
In the slow collapse model (c), CO is depleted over a larger area, so that
the gradient of the CO abundance is even shallower than in the L-P model.

The centrally peaked N$_2$H$^+$ abundance distribution in our models
is caused by the
relatively slow formation of N$_2$ in the outer regions, which are at
low density. The degree of CO depletion, on the other hand, is dependent on
its adsorption energy
(Paper I). To determine if the agreement with the results of
\cite{taf2002} can be improved, we performed new calculations
in which the nitrogen
is all in the form of N$_2$ initially and the adsorption energy of CO
on grains is varied. In our models discussed up to now, the
adsorption energy for CO is that of pure CO ice, which is 960 K.  The
CO adsorption energy is higher on other surfaces; e.g., it is 1210 K
if CO is adsorbed on silicate material and 1780 K on water (polar) ice
(Paper I and references therein).  We have run models with both 1210
K and 1780 K for the adsorption energy of CO.
The resulting distributions are shown in Figure \ref{fig:dist_L1544} (d-f).
  As can be seen, the steep drop in CO abundance is
well reproduced in the model with $f=3$ and an adsorption
energy of 1780 K (f).
%An alternative solution
%for the steep CO gradient is to form the central region of the core first, so
%that the depletion is heavy there {\sc Yuri - I don't fully understand},
%  and to form the outer region of the core
%later via accumulation of clumps, as discussed in \citet{klv1996} and
%Paper I.
Concerning the distribution of N$_2$H$^+$, the constant abundance observed
in L1544
is better reproduced with the N$_2$-rich initial abundance, although
the initial fraction of N$_2$ should be smaller than unity in order to
reproduce the absolute abundance. The improvement in the calculated
N$_2$H$^+$ distribution suggests either
that early formation of N$_2$ is more efficient than assumed in our
standard model, or that some molecular development occurs before
the collapse stage of prestellar cores.

How does our current agreement with observation in L1544 compare with
those based on previous models ?  In Paper I we preferred the Larson-Penston
model over models with slower collapse.  With this model, we were
able to best reproduce the radial distributions of the column
densities of CCS, CO, and N$_{2}$H$^{+}$ as well as the peak column
densities of CCS and CO.  Our calculated column density for N$_{2}$H$^{+}$
was low by a factor of 25, however.  The current results are a
considerable improvement since they show agreement over a wider group
of species, which includes N$_{2}$H$^{+}$.  This agreement is
obtained without slowing down the core collapse.
%consideration of a retarding magnetic field.
Moreover, we see no improvement when the collapse is retarded except for
fitting the results of Tafalla et al. (2002), which also requires some fraction
of nitrogen to be initially in its molecular form and CO to have a much higher
adsorption energy. The
improvement from Paper I is thus due to the inclusion of grain
chemistry.

\citet{li2002} include surface chemistry in their model.  Their
chemistry is not identical with ours because they use a small subset of the
UMIST gas-phase network and our grain surface reactions.
With their networks and the absence of a magnetic field,
they obtain reasonable agreement with the data except that CS and CCS
are found to be centrally peaked, in disagreement with observation and
also in disagreement with our results. Considering the results of \S 3.6,
the difference between our models could be caused by the use of
different chemical networks.
When a magnetically retarded collapse is
used, they find
%in general modest differences except that the
%distributions of CS and CCS do begin to show
a spatial hole at the core.
Their magnetic model is best compared with our model in which
$f=3$ and $S=0.33$ since in the work of \citet{li2002} the sticking
probability is assumed to be 0.3, and the collapse time scale of the
magnetized cloud is larger than that of non-magnetized cloud by a
factor of $\sim 2$ when the central density is $10^5-10^6$ cm$^{-3}$.
%Both models provide reasonable fits to the data, confirming that the
%inclusion of surface chemistry is vital, especially for N$_{2}$H$^{+}$
%and that the effective retardation factor $f$ must be $\lesssim 3$ when
%the central density of the core is $\sim 10^6$ cm$^{-3}$.
%\footnote{``Both...'' or should we say ``
Agreement between our model with $S=3$ and
$f=0.33$ and observation is improved from Paper I by the inclusion of
the grain-surface reactions, although it is certainly not our
best model.
Although some differences still remain between this
current model and \citet{li2002}, we can conclude that the main cause of the
difference between our Paper I and \citet{li2002} is grain-surface
chemistry.

\begin{deluxetable}{l c c c c c c}
\tabletypesize{\scriptsize}
\tablecaption{Molecular Column Densities (cm$^{-2}$) in L1544 and
Theoretical Models
\label{L1544}}
\tablewidth{0pt}
\tablehead{
\colhead{Species} & \colhead{L1544}   &
\colhead{f=1, S=1.0} & \colhead{f=3, S=1.0}  & \colhead{f=3, S=0.33} &
\colhead{f=10, S=1.0} & \colhead{f=10, S=0.1}}

\startdata
CO         & 1.2(18) \tablenotemark{a} &
{\bf 1.1(18)} & {\bf 6.1(17)} & {\bf 2.0(18)} & 3.3(17) & {\bf 3.0(18)} \\
CCS        & 2.6(13)\tablenotemark{b} &
{\bf 1.9(13)} & {\it 3.0(12)} & {\bf 9.6(12)} & {\it 2.0(11)} & 5.7(12) \\
N$_2$H$^+$ & 2.0(13)\tablenotemark{a} &
{\bf 1.3(13)} & {\bf 7.7(12)} & {\it 2.7(12)} & {\bf 7.5(12)} & {\it 3.3(12)}\\
N$_2$D$^+$ & 4.3(12)\tablenotemark{a} &
{\bf 5.1(12)} & {\bf 2.7(12)} & {\it 6.6(11)} & {\bf 2.0(12)} & {\it 4.8(11)}\\
NH$_3$     & 4.0(14)\tablenotemark{c} &
{\bf 3.6(14)} & {\bf 2.3(14)} & {\bf 2.2(14)} & {\bf 2.0(14)} & {\bf 3.6(14)}\\
CS         & 4.6(13)\tablenotemark{c} &
{\bf 4.7(13)} & 1.5(13) & {\bf 4.1(13)} & {\it 2.3(12)} & {\bf 2.8(13)}\\
HNC        & 1.5(14)\tablenotemark{d} &
{\bf 3.7(14)} & {\bf 1.6(14)} & {\bf 1.7(14)} & {\bf 8.6(13)} & {\bf 2.8(14)}\\
HCO$^+$    & 1.1(14)\tablenotemark{a} &
{\bf 1.3(14)} & {\bf 1.3(14)} & {\bf 1.8(14)} & {\bf 1.0(14)} & {\bf 1.7(14)}\\
DCO$^+$    & 4.0(12)\tablenotemark{a} &
1.8(13) & {\it 2.6(13)} & {\it 2.8(13)} & 1.7(13) & {\it 3.2(13)} \\
      \enddata

%% Text for table notes should follow after the \enddata but before
%% the \end{deluxetable}. Make sure there is at least one \tablenotemark
%% in the table for each \tablenotetext.

\tablenotetext{a}{\citet{cas2002b}}
\tablenotetext{b}{\citet{oha1999}}
\tablenotetext{c}{Calculated from best fit model of \citet{taf2002},
assuming a spherical core with radius of 15000 AU.}
\tablenotetext{d}{\citet{hir2003}}

%\tablecomments{Occasionally, authors wish to append a short
%paragraph of explanatory notes that pertain to the entire table, but
%which are different than the caption.  Such notes should be placed in
%a {\tt tablecomments} command like this.}

\end{deluxetable}
%\clearpage
\subsection{Variation among Cores}

Several groups have performed surveys of dense cores through
various molecular lines \citep{su1992,hir1998,taf2002,cas2002b}.
Comparison of the results of our model with these surveys should be
useful in investigating how molecular column densities vary during
the evolution of cores.  Table \ref{survey} lists observed column densities of
assorted species in 10 prestellar cores. We chose compact ($\sim 10^4$ AU)
prestellar cores, which, we believe, have just begun
collapse or are on the verge of star formation.
We carefully selected
the data obtained towards the peak position of N$_2$H$^+$ emission, in
order to avoid contamination with the spatial variation within the
cores.  The column densities in Table \ref{survey} with
parentheses are observed somewhat ($\sim 30''$) offset from the
N$_2$H$^+$ peak.  Since different beam sizes can also cause variations
in estimated column densities, we chose observations with similar beam
sizes for each molecular species.  If the core has been observed and
analyzed by \citet{taf2002}, we calculate from their best-fit model
the surface-averaged column density within the beam size of
\citet{cas2002c} ($54''$) for N$_2$H$^+$, \citet{su1992} ($80''$ and
$62''$) for NH$_3$ and CCS, and \citet{hir1998} ($34''$) for CS. The
column density of CO is averaged over the central radius of 2000 AU.
The molecular column densities of L1544 listed in Table 2 are slightly
different from those in Table 1, because of the larger beam sizes
adopted in Table 2.  Judging from the errors explicitly given for
N$_2$H$^+$ and differences in estimated column densities by
independent observers, the listed values may contain uncertainties by
a factor of $\sim 2-3$.  Large variations in the column densities of
certain species are still apparent among cores, especially among the
four objects at the top and bottom of Table 2.  Specifically,
in L1521E and L1521B, column densities of
N$_2$H$^+$ and NH$_3$ are smaller and those of CCS and CS are larger by
about an order of magnitude than in L63 and TMC-2A.

\begin{figure}
\epsscale{0.9}
\plotone{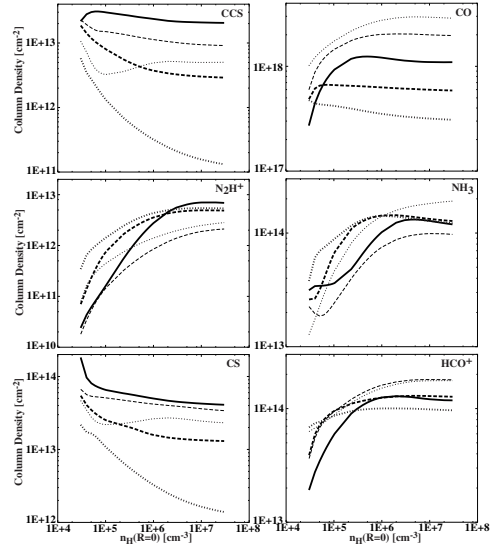}
\caption{Computed molecular column densities of assorted species
as a function
of the central density of the core.  Column densities are
surface-averaged over a radius of 4340 AU (which corresponds to 31$''$ assuming
the distance to the core to be 140 pc), 2000 AU, 3780 AU (27$''$), 5600 AU
(40$''$), 2380 AU (17$''$), and 2000 AU from the core center for CCS, CO,
N$_2$H$^+$, NH$_3$, CS, and HCO$^+$, respectively.
The solid lines represent the Larson-Penston core, while the dashed
and dotted lines represent more slowly collapsing cores by a factor
$f$ of 3 and 10, respectively.  The sticking probability is 1.0 for
thick lines, and $1/f$ for thin lines.\label{fig:den_avr}}
\end{figure}

Let us first consider how our results depend on the central density
($n_{\rm H}$) of the core.  Figure \ref{fig:den_avr} shows calculated
molecular column densities as a function of the central density in
our models when the central density is $n_{\rm H}\gtrsim 3\times 10^4$
cm$^{-3}$. The column densities of CCS, N$_2$H$^+$, NH$_3$, and CS are
surface-averaged
within the typical beam sizes mentioned above at the core center, while those
of CO and HCO$^+$ are surface-averaged within a radius of 2000 AU.
The column densities
of N$_2$H$^+$ and NH$_3$ tend to increase with increasing central
density; i.e., they increase as the core evolves.  The ion N$_2$H$^+$
is of special interest because its column density increases
monotonically in all of our models, including the models with an
initially high
abundance of N$_2$ discussed in \S 4.1.
In addition, the N$_2$H$^+$
column density does not depend significantly on the total mass (size)
of the core, because its absolute abundance is higher in the inner
regions. Hence this ion appears to be a useful probe of core evolution.
On the contrary, CCS and CS decrease with time in most of the models,
and CO and HCO$^+$ show little temporal variation except for the very early
stages, when $n_{\rm H}\lesssim 10^5$ cm$^{-3}$. A variation among 
core models is
apparent for CCS, CO, and CS.

In Table \ref{survey} we have arranged the objects in the order of
increasing N$_2$H$^+$ column density, assuming this order to represent
evolutionary stage.   We see that Table \ref{survey} indicates
that the column density of NH$_3$ tends to increase
with increasing time, which
is consistent with our model results, as is the decrease in CCS
and the weaker but similar trend for CS
(see Figure \ref{fig:den_avr}).  Another interesting fact is that
in L1521E and L1521B, which are considered to be the youngest ones in Table
\ref{survey} according to the N$_2$H$^+$ column density, CCS emission peaks at
the core center \citep{oha2000,hir2002}. Such a CCS distribution is indeed
characteristic of a very early stage of core evolution in our models.

Since the variation among core models is relatively significant for CCS,
CO, and CS, a quantitative comparison with observation
  might exclude some collapse models,
assuming that the collapse time scale (i.e. the parameter $f$) is the same
in all cores. For example, the computed
temporal variation of CCS is much smaller in the L-P model than the
observed variation
in Table \ref{survey}. However, the morphology of the core should be considered
in a quantitative comparison of these species, which are heavily 
depleted in the
central region.  Our caution arises because
the temporal variation of their model column densities is small at
least partially
because of the spherical symmetry assumed in our calculations. Although
their abundances decrease significantly in the central regions as the
core evolves (Figure \ref{fig:dist}), contributions from the outer
radii keep the molecular column densities almost constant.  In reality,
cores are not spherical, and depletion in the central region will be
more apparently reflected in the molecular column density.
Models of non-spherical cores are desirable for further studies.

%Although our model qualitatively explains the variation of molecular
%abundances
%among cores by core evolution, it fails in a quantitative comparison.
%While the column densities of CCS and CS vary by an order of magnitude
%in Table \ref{survey}, the temporal variation is very small in
%each model except for the model with $f=10$ and $S=1.0$, in which the
%absolute values of the CCS and CS column densities are much smaller than
%observed.
%The temporal variation of the CCS and CS column densities is small
%because of the spherical symmetry assumed in our model calculations.  Although
%their abundances decrease significantly in the central regions as the
%core evolves (Figure \ref{fig:dist}), contributions from the outer
%radii keep the molecular column densities almost constant.  In reality,
%cores are not spherical, and depletion in the central region will be
%more apparently reflected in the molecular column density.
%Models of non-spherical cores are desirable for further studies.

L1544 has significantly higher column densities of CCS and CO compared with
other objects with a similar N$_2$H$^+$ column density. There are a few
possible explanations. Firstly, the difference can be caused by core
geometry and/or
view angles. If the cores are not spherical and are observed with a
nearly face-on
angle, we would obtain significantly lower column densities of CCS and CO than
in the case of a spherical core or a non-spherical core with edge-on geometry,
as discussed above. For this explanation to be germane, L1544 would
be a spherical core or
  a core with
edge-on geometry, while the others would be non-spherical cores with a
more face-on geometry.
Interestingly, \citet{oha1999} suggested from their CCS observation that L1544
is a non-spherical core that is observed with a nearly edge-on geometry.
\citet{cb2000} expected a similar geometry by comparison of the
object with their magnetized core model.
%\footnote{added}\footnote{THIS IS A REAL FOOTNOTE: Agreement of our spherical
%core models with L1544 discussed in \S 4.1 does not contradict with the
%non-spherical and edge-on geometry suggested by \citet{oha1999}, as long as
%the density distribution in L1544 along the line of sight is similar to that
%in our spherical model.}
Another possible explanation is that the collapse time scale is different among
cores. While L1544 is best reproduced by the Larson-Penston core in our
model, other objects may collapse more slowly (e.g. by ambipolar diffusion).

%\clearpage
\begin{deluxetable}{l c c c c c}
\tabletypesize{\scriptsize}
\tablecaption{Observed Molecular Column Densities in Assorted  Prestellar Cores
\label{survey}}
\tablewidth{0pt}
\tablehead{
\colhead{Object} & \colhead{N$_2$H$^+$}   &
\colhead{NH$_3$} & \colhead{CCS}  & \colhead{CS} & \colhead{CO} \\
\colhead{} & \colhead{$10^{12}$ cm$^{-2}$}   &
\colhead{$10^{14}$ cm$^{-2}$} & \colhead{$10^{12}$ cm$^{-2}$} &
\colhead{$10^{14}$ cm$^{-2}$} & \colhead{$10^{17}$ cm$^{-2}$}\\}

\startdata
L1521E  & $<0.14$\tablenotemark{a} & 0.73\tablenotemark{a} &
     28\tablenotemark{a} & 3.0\tablenotemark{a} & \\
L1521B  & 1$\pm$0.85\tablenotemark{b}& 0.6\tablenotemark{c} &
     36\tablenotemark{c} & 1.3\tablenotemark{d} & \\
L1517B  & 3$\pm 0.3$\tablenotemark{e}, 3.1\tablenotemark{f} &
     7\tablenotemark{c}, 2.1\tablenotemark{f} &
     8.6\tablenotemark{c} & 0.13\tablenotemark{f} & 1.5\tablenotemark{f}  \\
%L1536   & 3.9$\pm 0.4$\tablenotemark{e} & 4.9\tablenotemark{c} &
%  6.6\tablenotemark{c} &     &   \\
L1400K  & 4$\pm 2$\tablenotemark{e}, 2.7\tablenotemark{f} &
     1.8\tablenotemark{c}, 7.2\tablenotemark{f} &
     3.6\tablenotemark{c} & 0.25\tablenotemark{f} & 2.1\tablenotemark{f}  \\
L1512   & 3$\pm$2\tablenotemark{e}  & (7)\tablenotemark{c,g} &
     (2.9)\tablenotemark{c} &  \\
%TMC-2   & 4.9$\pm 0.5$\tablenotemark{e} & 6.1\tablenotemark{c} &
%  25\tablenotemark{c} & 0.69\tablenotemark{d} &   \\
L1498   & 8$\pm$ 4\tablenotemark{e}, 3.0\tablenotemark{f} &
     (4.1)\tablenotemark{c},
     2.3\tablenotemark{f} &
     (16.5)\tablenotemark{c} & (0.40)\tablenotemark{d}, 0.14\tablenotemark{f} &
     2.4\tablenotemark{f} \\
L1495   & 6.0\tablenotemark{f}      & 2.4\tablenotemark{f} &
                             & 0.13\tablenotemark{f} & 3.6\tablenotemark{f}\\
L1544   & 9$\pm 2$\tablenotemark{e}, 7.3\tablenotemark{f} &
     1.8\tablenotemark{f} &
     20\tablenotemark{h} & 0.46\tablenotemark{f} & 4.0\tablenotemark{f},
12\tablenotemark{i} \\
L63     & 8$\pm 4$\tablenotemark{e} & 7.9\tablenotemark{c} &
     $<1.7$\tablenotemark{c} &  0.36\tablenotemark{d} & \\
TMC-2A  & 11$\pm 3$\tablenotemark{e} & 10.7\tablenotemark{c} &
     3.2\tablenotemark{c}    &       &  \\
     \enddata

%% Text for table notes should follow after the \enddata but before
%% the \end{deluxetable}. Make sure there is at least one \tablenotemark
%% in the table for each \tablenotetext.

\tablenotetext{a}{\citet{hir2002}}
\tablenotetext{b}{Takakuwa et al. in prep}
\tablenotetext{c}{\citet{su1992}}
\tablenotetext{d}{\citet{hir1998}, assuming $^{34}$S/S= 4.2\%}
\tablenotetext{e}{\citet{cas2002c}}
\tablenotetext{f}{\citet{taf2002}}
\tablenotetext{g}{The column densities with parentheses are observed
somewhat ($\sim 30''$) offset from the N$_2$H$^+$ peak.}
\tablenotetext{h}{\citet{oha1999}}
\tablenotetext{i}{\citet{cas2002b}}

%\tablecomments{Occasionally, authors wish to append a short
%paragraph of explanatory notes that pertain to the entire table, but
%which are different than the caption.  Such notes should be placed in
%a {\tt tablecomments} command like this.}

\end{deluxetable}
%\clearpage
\subsection{D/H Ratios In Assorted Cores}
Finally, we compare the molecular D/H ratios obtained in our models
with observations in a variety of cores.  As mentioned in \S 3.7,
molecular D/H ratios are predicted to increase during the evolution of
the cores.  It would be
useful to plot these ratios vs the central densities, which are a
direct measure of the evolution of the cores.  Since we do not know
the central density (i.e. evolutionary stage) of most evolving cores,
we can use the column density of the ion N$_{2}$H$^{+}$ as an
indicator of evolution, following our previous arguments in \S 4.2.
This ion has been studied in approximately 60 cores by
\citet{cas2002c}, while the column density ratio DCO$^+$/HCO$^+$ has
been measured
by \citet{bu1995} and \citet{wil1998}, and the ratio DNC/HNC has been
measured by \citet{hir2001}
and \citet{hir2003}, both towards several prestellar cores.
The ratio of N$_{2}$D$^{+}$/N$_2$H$^+$ in L1544 has been studied by
\citet{cas2002b}.
We emphasize that the column
density of N$_2$H$^+$ does not depend significantly on the total mass
of the core, because of the centrally peaked distribution of its absolute
abundance.  The D/H
column density ratios are not significantly dependent on the total
mass either, to the extent that the dependence is at least smaller
than that of the molecular column densities themselves.  Thus, it is
appropriate to include cores of differing mass on the same
evolutionary plot.

\begin{figure}
\plotone{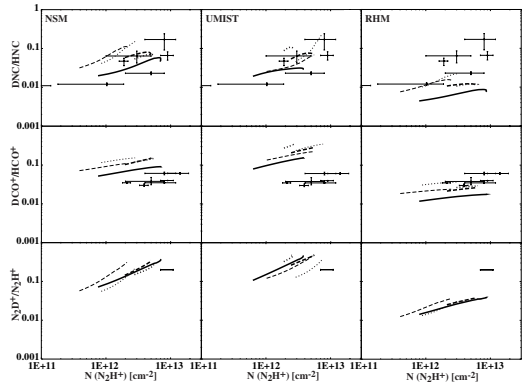}
\caption{Column density ratios of deuterated to normal species as a
function of suitably averaged N$_2$H$^+$ column density.  The solid
lines represent the Larson-Penston core, while the dashed and dotted
lines represent more slowly collapsing cores by a factor $f$ of 3 and
10, respectively.  The sticking probability is 1.0 for thick lines,
and $1/f$ for thin lines.  The observational results are depicted with
error bars; horizontal and vertical lines indicate uncertainties in the
N$_2$H$^+$ column density and molecular D/H ratio, respectively.
\label{fig:N2Hp_ratio}}
\end{figure}

Figure \ref{fig:N2Hp_ratio} contains our plot of D/H column density
ratios for HNC, HCO$^{+}$ and N$_2$H$^+$ versus the N$_2$H$^+$ column
density. Again we chose the observational data which are measured at or close
to the emission peak of N$_2$H$^+$: L1498, L134A, L1544, L1696A, L183(S), L63,
and L1155C for the plot of DCO$^+$/HCO$^+$, and L1521B, L1082A, L1521E, L1512,
L1544, L63, and L1155C for DNC/HNC.
Figure \ref{fig:N2Hp_ratio} shows that D/H ratios tend to
increase with N$_2$H$^+$ column density, which is in qualitative agreement
with the model results.
    For each molecule, the results from three different
gas-phase chemical networks are depicted in order to consider
uncertainties arising from uncertainties in the chemistry.  In
addition to the NSM and UMIST networks discussed earlier, we utilize
the NSM network with the new rates for deuterium fractionation of
\citet{ghr2002} as utilized by \citet{rhm2002}, which we label RHM.
The theoretical results, plotted as lines, encompass our model cores
as their central density increases from $3\times 10^5$ cm$^{-3}$ to
$3\times 10^7$ cm$^{-3}$.  The calculated molecular column densities
are averaged over radii of 1400 AU, 4900 AU, and 3780 AU from
the core centers for HNC (DNC), HCO$^+$ (DCO$^+$), and N$_2$H$^+$,
respectively, considering the beam sizes of the observations.
In calculating the ratio of N$_2$D$^+$/N$_2$H$^+$, we averaged the column
density over a radius of 1400 AU, referring to the beam size of
\citet{cas2002b}.  For each D/H ratio and model network, we calculated
results for the Larson-Penston case, cases with collapse delayed by
factors $f$ of 3 and 10, and cases in which a sticking probability of
$1/f$ is assumed.
%\footnote{``With the exception of L63, which is located in $\rho$ Ophiuchus,
%all other observed cores lie in the Taurus molecular cloud.'' is omitted here,
%because new plot contains sources in other regions. Should we list the object
%name in the text or in the figure ? One problem is that DNC/HNC by Hirota
%et al. is in preparation. He knows that we are going to use his data.
%But I am not sure listing the source names is OK for him. YURI: IF WE
%ARE TALKING OF A FEW CORES IN THE FIGURES, I WOULD NAME THEM IN THE
%PREVIOUS SECTION. -- Object names are added at the top of this paragraph.}
As discussed,
the theoretical plots tend to show a direct relation between the
D/H ratios and the N$_2$H$^+$ column density.
The ratio of DNC/HNC is of special interest
because its dependence on the collapse model is relatively
significant.
%Only for the case of
%DNC/HNC is there enough observational evidence to buttress this
%theoretical prediction.

If we initially restrict our attention to the NSM calculations, we see
that the standard Larson-Penston solution is the best overall
representation of all but the largest DNC/HNC value, which occurs for
L63.  This agreement, however, is not maintained with the RHM
network, in which less efficient fractionation must be balanced by
the adoption of very slow collapse models.  Even so, the RHM model
cannot account for the high N$_2$D$^+$/N$_2$H$^+$ ratio observed
in L1544.  The results with the UMIST network also vary somewhat from
those of the NSM network.  In summary, we can state that despite all
of the uncertainty, the use of the new RHM values makes accounting for
deuterium fractionation in a quantitative manner far more difficult, a
result previously obtained by \citet{rhm2002} for static cores.

\section{Summary}

An epoch in low-mass star formation prior to the well-known Class 0
stage is now well established.  In this prior stage, which is now
labeled a prestellar core, there is evidence for
collapse towards a central condensation although this condensation is
still at the low temperature of the cloud.  The observation of
molecular column densities and radial
distributions of these column densities in  prestellar cores
yields information on the nature of the collapse if it can be properly
interpreted.  For such an interpretation, a detailed understanding of
the chemical processes is needed.  In our first paper on the subject
(Paper I), we modeled the gas-phase chemistry occurring during
collapse according to the Larson-Penston solution.  We found that
both the average molecular column densities and their radial
distributions  for CCS, CO, and N$_{2}$H$^{+}$ in the prestellar
core L1544  were only
partially reproduced by the model.  The most salient disagreement occurred
in the column density of the molecular ion N$_{2}$H$^{+}$, which was
determined to be
much lower than observed. Slower collapse models did not improve the
situation; N$_2$H$^+$ is enhanced but CCS column densities become
significantly lower than observed.
More recently, \citet{li2002} performed similar
calculations with two important additions: a small number of chemical
reactions occurring on grain surfaces and a magnetic field to retard
collapse.  They found better agreement with observations in L1544,
although it remained somewhat unclear whether the source of the
improvement was the inclusion of surface chemistry or the magnetic
field, or both.

In the present paper, we have included a complete set of surface
chemical reactions into our chemical network and redone the
calculations.  There is now much more observational information with
which to compare results - both in L1544 and in other condensations.
%{\sc some of which are prestellar while others are often described as
%`starless' rather than 'prestellar' because of their relatively low density.}
We find that our improved Larson-Penston model is now able to
reproduce most  of the observational data, especially in L1544.
Furthermore, our current model results are in better agreement with
\citet{li2002}
than shown by a comparison between \citet{li2002} and our Paper I; ammonia
and N$_2$H$^+$ are more abundant, and the dependence of
CCS column density on collapse time scale is smaller in the current model.
     The radial distributions of molecules are in reasonable agreement
with those of
\citet{li2002} for their magnetized core, whether or not we use our
standard case or slow the collapse down somewhat to mimic the
magnetized result.  Interestingly, the
agreement is not as good with their unmagnetized core. Specifically, their
radial column density distributions of CS and CCS are strongly peaked
at the core center in the non-magnetic case whereas ours are not.
One must remember, in this comparison, that they use a subset
of the UMIST
network and we use the NSM network; differences in the abundances of
sulfur-bearing species between the two networks are well known.
On the whole, however, we may conclude
that the inclusion of surface chemistry, however uncertain it
      may be, is clearly necessary.

     How good is the current state of agreement between our model and
     observational results?  Restricting attention to L1544, we find that
     both the radial column density distributions of the
     well-studied species CS, CCS,
     N$_{2}$H$^{+}$ and CO and averaged column densities of 9
     species are in the main
     reasonably fit by a variety of models, including the standard
     Larson-Penston case and variations with a small degree of delayed
     collapse and/or reduced sticking efficiencies of gas-phase molecules
     on grains.   Specifically, our models can reproduce averaged column
     densities of up to eight species to within a factor of three.
     When the collapse is delayed by a factor of ten, however,
     the agreement is less good.
     The inferred radial distributions expressed in terms of fractional
     abundances are somewhat harder to reproduce with our standard
     initial conditions and an adsorption energy for CO
     based on the value for pure CO ice.  A Moderately slow collapse model
     ($f\sim 3$) in which at least some
     of the nitrogen is initially in its molecular form and the
     adsorption energy of CO is increased significantly to its value
     for pure water ice is much more successful. 

%    The molecules in good agreement with
%    observation include the deuterated species
%%DCO$^{+}$ and
%    N$_{2}$D$^{+}$, showing that our model reasonably reproduces the
%    deuterium fractionation occurring.  This agreement is worsened,
%    however, when new results for selected deuteron/proton exchange
%    reactions are used \citep{ghr2002}.  {\sc Yuri: comparing this
%    last discussion with that in the second paragraph below, you seem to be
%    making a distinction between the absolute abundance of the
%    deuterated ion and the D/H ratio.  Yet your analysis - high
%    depletion - is the same, as is the discussion of the negative
%    effect of the newly measured reaction rates.  Do you think that we
%    need both discussions?  There does seem to be repetition.}

%  If we consider a larger sample of prestellar cores, the situation is
%  less clear.
We have also attempted to consider a larger sample of prestellar cores
%mainly {\bf starless condensations}, and
and
see how our models can help to understand the varying degrees of
evolution of the cores using two primary indicators of evolution: the
central density of the core, and the central column density of the
ion N$_{2}$H$^{+}$, which is found theoretically to increase
monotonically with increasing core density.  With these indicators, we
are able to qualitatively reproduce the observed dependence of most of
the molecular column densities.  The column density of NH$_3$ tends to
increase, and those of CCS and CS to decrease with core evolution.
The non-spherical geometry of cores should also be taken into account,
however, for more quantitative comparisons of species which are heavily
depleted in the central regions.
%such as distinguishing
%collapse models based on molecular column densities.

     Another good indicator of evolution is predicted to be the D/H ratio
     of molecular species, since fractionation is enhanced when density
     increases and heavy species are adsorbed onto grain surfaces.
     Indeed a plot of the observational data shows that the ratios of DNC/HNC
     and DCO$^+$/HCO$^+$ increase with the N$_2$H$^+$ column density.
%  Only for the ratio DNC/HNC, however, does there appear to be
%  sufficient observational information to see an evolutionary
%  dependence.
     The dependence of DNC/HNC on the N$_2$H$^+$ column density is in the
     main reproduced well by our standard Larson-Penston core and slight
     variations in collapse rate.
     The column density of N$_2$D$^+$ and the high column density ratio of
     N$_2$D$^+$/N$_2$H$^+$ observed in L1544 are also reproduced in our
     models.
     There is one major problem
     though: the extended NSM network used in our calculations does not
     contain some new rate coefficients measured by \citet{ghr2002}; the
     effect of including these rates, which remain to be confirmed by
     other experiments, is to lower the fractionation to levels
     unacceptable unless very slow collapse is assumed.  Even so, the
     agreement is not as good as for the plain Larson-Penston case with
     the NSM network.  This puzzling disagreement mimics that found by
     \citet{rhm2002} for static cores.  It is clear that deuterium
     fractionation is not yet completely understood.

As more prestellar cores are observed in greater degrees of
detail, it should be possible to distinguish among chemical treatments better
and to use model calculations to better constrain the evolutionary dynamics.
Modeling of non-spherical cores and direct coupling of the large chemical
network, (magneto-)hydrodynamic calculation of cores and radiation transfer
are desirable
for comparison with observed line profiles and intensity maps.
Since surface chemistry appears to play an important role,
improvements in the currently quite approximate treatment of surface
processes would appear to be a necessity.  Such improvements are
currently being developed by several groups including our own.

\acknowledgments
We are grateful to Drs. S. Takakuwa and T. Hirota for providing their
observational data prior to  publication and for helpful discussions.
We appreciate constructive comments on a prior version of this
manuscript from both the referee (M. Tafalla) and Paola Caselli.
Y. A. is supported by a Grant-in-Aid for Scientific Research of the Ministry
of Education, Culture, Sports, Science and Technology of Japan
(13011203 and 14740130).
N. O. is supported in part by NSC grant
91-2112-M-001-029. The Astrochemistry program at The Ohio State
University is supported by the National Science Foundation.  Numerical
calculations were carried out on the VPP5000 at the Astronomical Data
Analysis Center of the National Astronomical Observatory of Japan.

\clearpage

%% Use the figure environment and \plotone or \plottwo to include
%% figures and captions in your electronic submission.

\end{document}